\documentclass[a4paper,fleqn]{cas-sc}

\usepackage[authoryear,longnamesfirst]{natbib}

\def\tsc#1{\csdef{#1}{\textsc{\lowercase{#1}}\xspace}}
\tsc{WGM}
\tsc{QE}

\begin{document}
\let\WriteBookmarks\relax
\def\floatpagepagefraction{1}
\def\textpagefraction{.001}

\shorttitle{q3-MuPa for Fast Multi-Parametric MRI}

\shortauthors{Wang et al.}  

\title [mode = title]{q3-MuPa: Quick, Quiet, Quantitative Multi-Parametric MRI using Physics-Informed Diffusion Models}  

\author[1]{Shishuai Wang}[orcid=0009-0008-8550-6591]
\cormark[1]
\ead{s.wang@erasmusmc.nl}
\credit{Conceptualization, Methodology, Software, Formal analysis, Investigation, Data curation, Visualization, Validation, Writing - original draft, Writing - review \& editing}

\author[2]{Florian Wiesinger}
\credit{Data curation, Resources, Project administration, Supervision, Writing - review \& editing}

\author[1]{Noemi Sgambelluri}
\credit{Data curation}

\author[2]{Carolin Pirkl}
\credit{Resources, Supervision, Writing - review \& editing}

\author[1]{Stefan Klein}
\credit{Methodology, Funding acquisition, Resources, Project administration, Supervision, Writing - review \& editing}

\author[1]{Juan A. Hernandez-Tamames}
\credit{Funding acquisition, Resources, Project administration, Supervision, Writing - review \& editing}

\author[1]{Dirk H.J. Poot}
\credit{Conceptualization, Methodology, Software, Investigation, Data curation, Funding acquisition, Resources, Project administration, Supervision, Writing - review \& editing}

\affiliation[1]{organization={Erasmus University Medical Center},
            addressline={Dr. Molewaterplein 40}, 
            city={Rotterdam},
            postcode={3015 GD}, 
            country={The Netherlands}}

\affiliation[2]{organization={GE HealthCare},
            addressline={Oskar-Schlemmer-Straße 11}, 
            city={Munich},
            postcode={80807}, 
            country={Germany}}

\cortext[1]{Corresponding author}



\begin{abstract}
MuPa-ZTE is a novel multi-parametric quantitative MRI protocol that enables fast and nearly silent scanning. 
However, accelerating multi-parametric acquisitions to meet clinical time constraints makes the reconstruction of accurate and clean qMRI maps increasingly challenging, particularly under severe undersampling and noise. 
In this work, we propose a physics-informed diffusion model for MuPa-ZTE qMRI mapping.
A denoising diffusion probabilistic model is trained to map MuPa-ZTE weighted image series to T1, T2, and proton density maps, while the MuPa-ZTE forward model is incorporated as an explicit data consistency constraint during inference. 
The proposed method is trained entirely on synthetic data and evaluated on both synthetic data and real data under the nominal ($\sim$4min) and fourfold-accelerated ($\sim$1min) MuPa-ZTE acquisitions.
Compared with dictionary matching and a purely data-driven diffusion model, the proposed approach yields accurate and less noisy 3D qMRI maps with improved structural fidelity.
The integration of MuPa-ZTE acquisition with a physics-informed diffusion model, termed q3-MuPa, provides an acquisition-consistent framework for quick, quiet, and quantitative multi-parametric MRI.
\end{abstract}


\begin{highlights}
\item Fast and nearly silent multi-parametric MRI using MuPa-ZTE
\item Physics-informed diffusion improves robustness of qMRI mapping
\item Accurate T1, T2 and PD estimation under fourfold-accelerated acquisition
\item Improved structural fidelity compared with conventional methods
\item Trained on synthetic data with strong generalisation to real data
\end{highlights}


\maketitle

\section{Introduction}
\label{introduction}
Magnetic Resonance Imaging (MRI) is a widely used non-invasive imaging technique that provides superior soft-tissue contrast and detailed anatomical or functional information. However, conventional MRI produces weighted images whose voxel intensities have no standardised physical meaning. 
In contrast, quantitative MRI (qMRI) aims to measure intrinsic tissue properties, such as T1 and T2 relaxation times and proton density.
These quantitative measurements enable objective comparison across scans and acquisition conditions and have attracted increasing interest as reproducible imaging biomarkers, as well as for synthesising arbitrary contrast-weighted images.

Despite this promise, the widespread adoption of qMRI in routine imaging remains limited by the difficulty of obtaining multiple quantitative parameters within clinically practical scan times. 
Estimating multiple parameters typically requires multiple contrast encodings or acquisitions, which can prolong examinations, increase sensitivity to motion, and complicate clinical workflows. 
Multi-parametric acquisition strategies have therefore been developed to jointly estimate multiple parameters within a single acquisition framework \cite{symri, mrf, 3dqalas, qti}.
Among these approaches, a novel 3D fast silent multi-parametric mapping sequence with zero echo time (MuPa-ZTE) has been proposed \cite{wiesinger20213d}. 
By employing a nominal zero echo time together with 3D radial phyllotaxis readouts \cite{piccini2011spiral, ljungberg2022motion}, MuPa-ZTE minimises gradient switching noise, enabling nearly silent scanning while improving motion robustness. 
Importantly, its phyllotaxis sampling scheme allows flexible acceleration of the acquisition, e.g., from nominal ($\sim$4min) to fourfold-accelerated ($\sim$1min), bringing the acquisition time closer to clinically practical ranges.
These properties make MuPa-ZTE a promising candidate for multi-parametric acquisition.

The baseline qMRI mapping strategy for MuPa-ZTE reconstructs a series of weighted images with varying contrasts and subsequently estimates quantitative parameters using dictionary matching based on the theoretical forward model \cite{wiesinger20213d, wiesinger2020psst}. 
While this model-driven approach is straightforward, the accuracy and visual quality of the resulting qMRI maps are often limited by noise, imaging artefacts, and the discrete resolution of the dictionary. 
When considering further acceleration of acquisitions, the quality of the subsequent qMRI mapping increasingly becomes a limiting factor in practice.

To improve qMRI mapping, learning-based approaches have been explored across a range of qMRI acquisition schemes. 
Purely data-driven convolutional neural networks (CNNs) have been proposed to learn the mapping between weighted image series and quantitative parameters \cite{cai2018single, jeelani2020myocardial, shao2020fast}, but their black-box nature and limited robustness outside the training distribution restrict their reliability for qMRI mapping.
Recent work incorporated physical signal models into learning-based reconstruction, such as recurrent inference machines (RIMs) that inject signal-model gradients during iterative inference \cite{sabidussi2021recurrent, zhang2022unified, sabidussi2023dtirim}.
More recently, diffusion models have emerged as powerful generative approaches and have also been applied to qMRI mapping. 
A purely data-driven denoising diffusion probabilistic model (DDPM) demonstrated favourable performance compared to a RIM in inversion-recovery acquisitions for T1 mapping \cite{wang2024qmri}, suggesting the potential of diffusion models for qMRI mapping. 
Other studies have incorporated physics information into diffusion frameworks \cite{bian2024diffusion, mayo2025physics}, reporting improvements in both visual quality and quantitative accuracy. 
However, for MuPa-ZTE in particular, these physics-informed diffusion-based methods require the modelling of k-space sampling, which makes the implementation challenging, owing to the substantial computational resources required for 3D processing and MuPa-ZTE's readouts at different contrast states. 
Consequently, a mapping approach that combines generative modelling with acquisition-consistent physics constraints while preserving computational feasibility for MuPa-ZTE remains lacking.

To address these challenges, we propose q3-MuPa, a qMRI framework that combines the MuPa-ZTE acquisition with a physics-informed diffusion model for qMRI mapping. 
To this end, we first train a DDPM to learn the distribution of qMRI maps conditioned on MuPa-ZTE weighted image series. 
During inference, data consistency (DC) is explicitly enforced by constraining the generated qMRI maps to be consistent with the acquired weighted images according to the MuPa-ZTE forward model. 
By incorporating physics constraints at inference without embedding full k-space processing into the diffusion pipeline, the proposed approach maintains agreement with the measured data while remaining computationally feasible for 3D MuPa-ZTE acquisitions. 
The proposed mapping method is evaluated against dictionary matching and a purely data-driven diffusion model.
To address the limited availability of training data, we generate synthetic MuPa-ZTE data. 
The diffusion models are trained entirely on synthetic data and validated on a NIST/ISMRM phantom, two healthy volunteers, and a patient with brain metastases.
We also assess performance under fourfold acceleration that enables qMRI mapping from approximately one minute acquisition.
In summary, this work contributes an acquisition-consistent diffusion-based reconstruction strategy tailored to MuPa-ZTE, addressing the emerging reconstruction limiting factor in fast multi-parametric imaging while preserving physical interpretability and computational practicality.

\section{Method}
\label{method}
\subsection{MuPa-ZTE Acquisition and Forward Modeling}
\label{dlmupa}

\begin{figure}
    \centering
    \includegraphics[width=\linewidth]{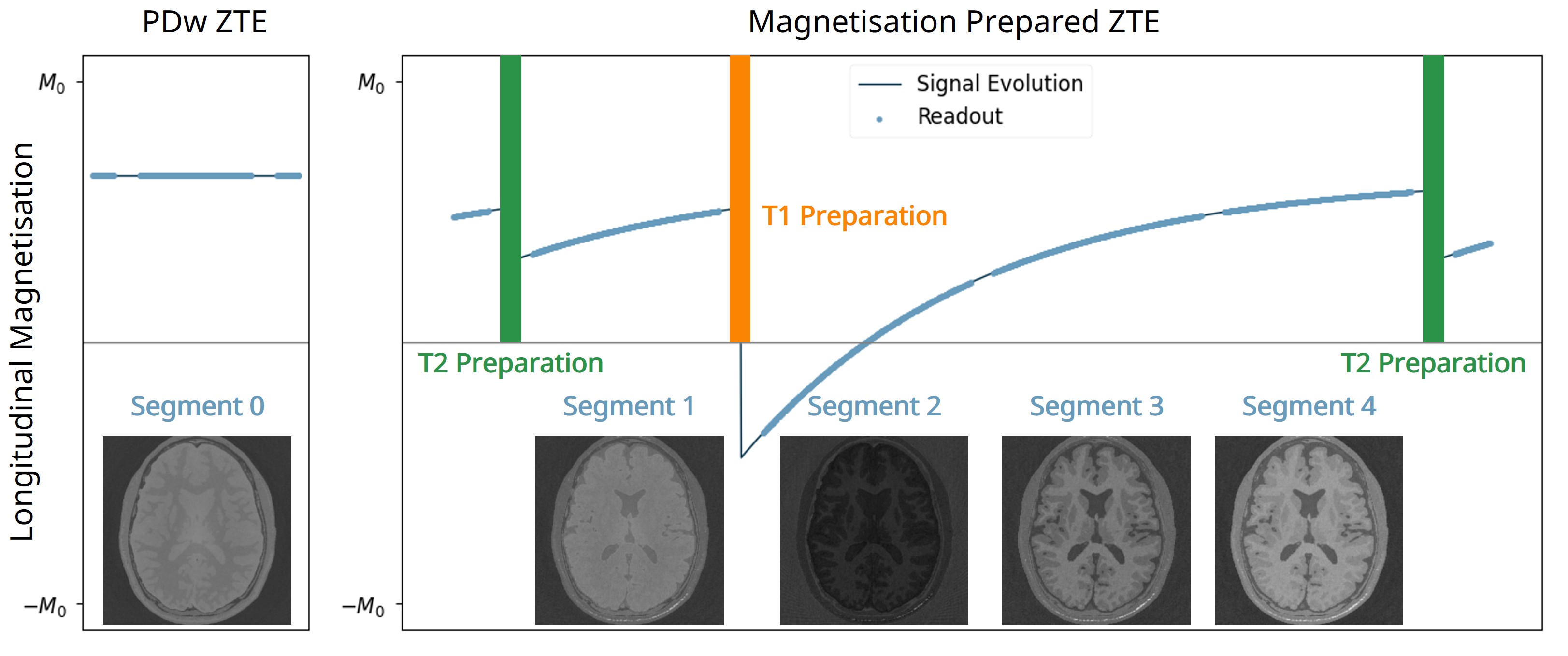}
    \caption{Schematic of a MuPa-ZTE acquisition. The weighted images at the bottom illustrate example contrasts for each segment.}  
    \label{fig1}
\end{figure}

MuPa-ZTE acquires five segmented weighted images (indexed by $s=0,\ldots,4$) with varying PD-, T1-, T2-contrasts (Fig.~\ref{fig1}).
A native ZTE image ($s=0$) is acquired upfront with a low flip angle ($\mathrm{FA_0}$) and its contrast can be described by the steady-state signal equation
\begin{equation}
\label{Mzspgr}
    M_{z,0}=PD \cdot \frac{1-E_1}{1-E_1 \cdot \cos \left(\mathrm{FA_0}\right)} \approx \frac{PD}{1+\frac{T_1}{TR} \cdot \frac{{\mathrm{FA_0}}^2}{2}}
\end{equation}
where $E_1=e^{-TR / T_1}$, $TR$ is the repetition time, and the approximation hold for $TR \ll T_1$  and $\mathrm{FA_0} \ll$ 1 rad \cite{ljungberg2021silent}.
The approximation indicates that $M_{z,0}$ is primarily PD-weighted with only mild T1 saturation.
Subsequently, four additional images are acquired with magnetisation-prepared ZTE readouts to impart T1 or T2 contrast.
The first and second magnetisation-prepared segments begin with a preparation pulse (T2-preparation and inversion, respectively), and each of the four segments is acquired using a train of ZTE readouts following excitation pulses with flip angle $\mathrm{FA}$.
Within a segment, each readout is performed at a different magnetisation state.
The longitudinal magnetisation right before the $i$-th readout in a segment $s$ is given by \cite{ljungberg2021silent}
\begin{equation}
\label{Mzi}
    M_{z,s,i}=M_{s,prep} \cdot \omega^i+M_{z, SPGR} \cdot\left(1-\omega^i\right)
\end{equation}
where $s=1,...,4$, $\omega=E_1\cos \left(\mathrm{FA}\right)$, $M_{z, SPGR}=PD \cdot \frac{1-E_1}{1-\omega}$ is the SPGR steady state, and $M_{s,prep}$ is the longitudinal magnetisation right before the first excitation pulse of the $s$-th segment (reflecting T1 or T2 relaxation effects).
Essentially, during the readout train the magnetisation relaxes from $M_{s,prep}$ towards the equilibrium determined by $M_{z, SPGR}$.
The resulting image contrast of a full magnetisation-prepared segment of $N$ readouts is the average magnetisation over the segment \cite{ljungberg2021silent, ljungberg2020silent}
\begin{equation}
\label{Mzseg}
    M_{z,s}=M_{s,prep} \cdot f(N, \omega)+M_{z, SPGR} \cdot[1-f(N, \omega)]
\end{equation}
where $f=\frac{1}{N} \cdot \frac{1-\omega^N}{(1-\omega)}$.
Equations~\ref{Mzspgr} and~\ref{Mzi} describe the evolution of the longitudinal magnetisation under the MuPa-ZTE preparation scheme.
The transverse magnetisation is obtained by projection onto the transverse plane via the corresponding flip angle, i.e., by multiplication with $\sin(\mathrm{FA_0})$ or $\sin(\mathrm{FA})$.
This transverse magnetisation is subsequently used to generate synthetic weighted images for training and evaluation of the diffusion models.
Similarly, Equations~\ref{Mzspgr} and~\ref{Mzseg}, together with the flip-angle projection, define the simplified MuPa-ZTE forward model, denoted as $\mathcal{F}(\cdot)$, which is used in both dictionary matching and the data consistency module during inference.

\subsection{Proposed Mapping Method}
\label{method_ours}
\begin{figure}
    \centering
    \includegraphics[width=\linewidth]{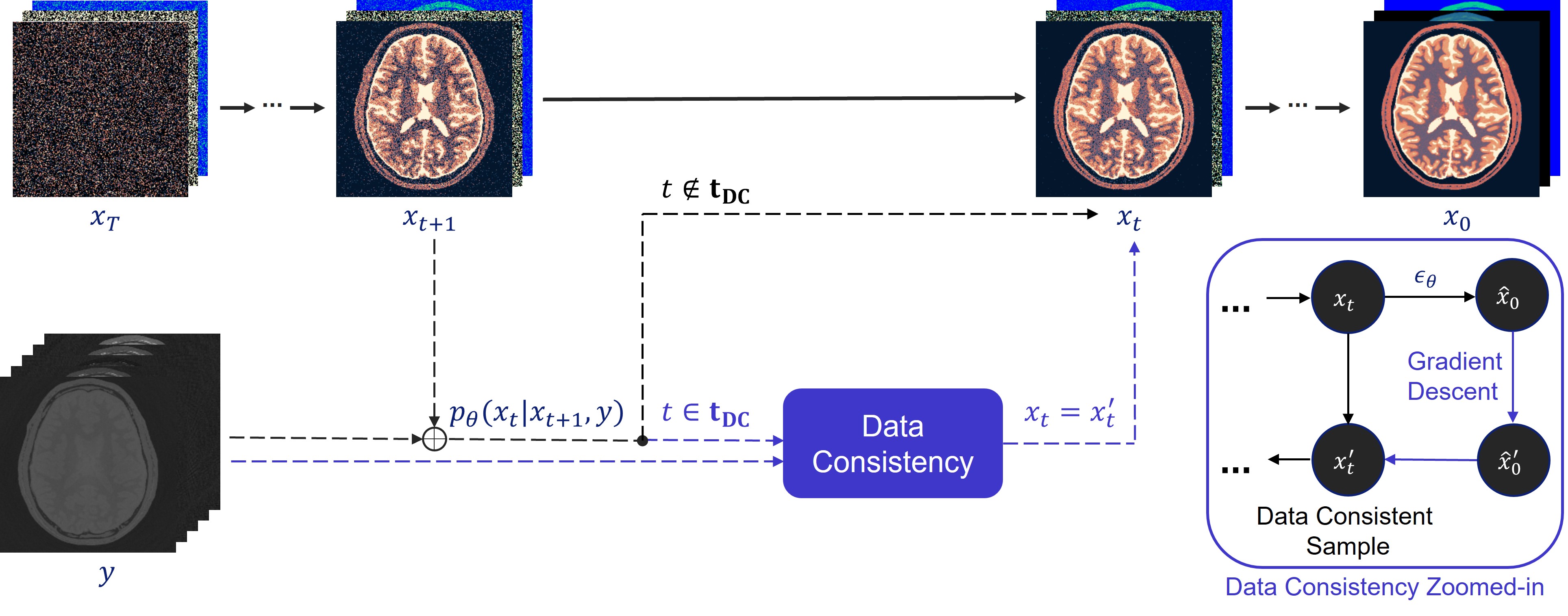}
    \caption{Schematic overview of the proposed inference workflow. 
    Starting from random noise, qMRI maps are generated through the reverse diffusion process conditioned on the input MuPa-ZTE weighted images ($y$). 
    At selected diffusion steps $\mathbf{t_{DC}}$, explicit data consistency is enforced by projecting intermediate estimates toward consistency with the MuPa-ZTE forward model via gradient-based optimisation.}  
    \label{fig2}
\end{figure}

We propose a diffusion-based qMRI mapping framework that generates qMRI maps from MuPa-ZTE weighted images while enforcing consistency with the MuPa-ZTE forward model. 
An overview of the inference workflow is shown in Fig.~\ref{fig2}.

\subsubsection{Diffusion Model for qMRI mapping}
Denoising diffusion probabilistic models (DDPMs) \cite{ddpm} learn data distributions through a progressive noising and denoising process. 
Given a clean sample $x_0$, the forward diffusion process progressively adds Gaussian noise over $T$ steps,
\begin{equation}
    \label{diffusion_fwd}
    q(x_{1:T} \mid x_0) = \prod_{t=1}^T q(x_t \mid x_{t-1})
\end{equation}
with
\begin{equation}
    \label{diffusion_fwd2}
     q(x_t \mid x_{t-1}):=\mathcal{N}(\sqrt{1-\beta_t}x_{t-1}, \beta_t\mathbf{I})
\end{equation}
where $\beta_1,...,\beta_T$ denotes a predefined variance schedule.
A model $p_\theta$ is trained to reverse this process by predicting the added noise conditioned on the noisy sample $x_t$, time step $t$, and measurements $y$. 
The reverse transition can be expressed as
\begin{equation}
    p_\theta\left(x_{t-1} \mid x_t, y\right):=\mathcal{N}\left(\mu_\theta\left(x_t, y, t\right), \Sigma\right)
\end{equation}
where the mean is parameterised using the predicted noise $\epsilon_\theta(x_t, y, t)$ \cite{ddpm}
\begin{equation}
    \mu_\theta(x_t, y, t) = \frac{1}{\sqrt{\alpha_t}}\left(x_t-\frac{\beta_t}{\sqrt{1-\bar{\alpha}_t}} \epsilon_\theta\left(x_t, y, t\right)\right)
    \label{mu_theta}
\end{equation}
with $\alpha_t:=1-\beta_t$ and $\bar{\alpha}_t:=\prod_{s=1}^t \alpha_s$.
The model is trained using the noise prediction objective
\begin{equation}
    L=\mathbb{E}_{t, x_0, \epsilon}\left[\left\|\epsilon-\epsilon_\theta\left(x_t, y, t\right)\right\|^2\right]
    \label{ddpm_loss}
\end{equation}

In our application, the clean sample $x_0$ corresponds to the stacked qMRI parameter maps (PD, T1, T2), while the five MuPa-ZTE weighted images serve as conditional inputs $y$ \cite{wang2024qmri}. 
Conditioning is implemented by concatenating $y$ with the noisy samples $x_t$ at the network input, enabling the denoising process to be guided by the acquired measurements.
After training, qMRI mapping is performed by starting from Gaussian noise $x_T$ and iteratively sampling from $p_\theta(x_{t-1}\mid x_t, y)$ until $x_0$ is obtained, yielding quantitative maps conditioned on the input images.
This formulation constitutes a purely data-driven conditional DDPM baseline.

\subsubsection{Data Consistency Integration}
The purely data-driven formulation enforces consistency with the input images $y$ only implicitly through conditioning, which may lead to physically inconsistent parameter estimates and increased stochastic variability.
To address these limitations, we integrate explicit data consistency into the reverse diffusion process following \cite{song2023solving}.
The key idea is to periodically project intermediate diffusion estimates toward consistency with the MuPa-ZTE forward model $\mathcal{F}(\cdot)$.
Unlike approaches that enforce data consistency directly in k-space, we operate in the weighted-image domain, since the reconstructed weighted images constitute the effective measurements used for parameter estimation in MuPa-ZTE mapping, consistent with the baseline dictionary matching.
Explicit k-space modelling would substantially increase computational complexity for 3D multi-contrast acquisitions while providing limited additional constraints for the mapping task. 

Specifically, at certain diffusion steps we pause the normal denoising and perform the following data consistency update:
\begin{enumerate}
    \item \textbf{Estimate the original sample.} At time step $t$, we compute a preliminary estimate $\hat{x}_0$ from the predicted noise \cite{ddpm}
    \begin{equation}
        \hat{x}_0=(x_t-\sqrt{1-\bar{\alpha}_t}\epsilon_\theta(x_t,y,t)) / \sqrt{\bar{\alpha}_t}
        \label{direct_estimation_x0}
    \end{equation}
    \item \textbf{Physics-constrained optimisation.} We obtain a corrected estimate $\hat{x}_0^{\prime}$ by enforcing consistency with the measurements through
    \begin{equation}
        \hat{x}_0^{\prime} \in \operatorname*{argmin}_{\hat{x}_0} \| \mathcal{F}(\hat{x}_0) - y \|_2^2
        \label{GD}
    \end{equation}
    where $\mathcal{F}(\cdot)$ denotes the MuPa-ZTE forward model.
    The optimisation is solved using gradient descent with early stopping. 
    Iterations terminate either when the optimisation loss falls below a predefined threshold $\tau$ or when a maximum number of iterations is reached, balancing measurement consistency against overfitting to noise or artefacts not explained by $\mathcal{F}(\cdot)$. 
    \item \textbf{Resample a consistent noisy state.} 
    The corrected estimate is mapped back to the reverse diffusion process by sampling a data-consistent noisy state $x_t^{\prime}$ 
    \begin{equation}
        p\left(x_t^{\prime} \mid x_t, {\hat{x}_0^{\prime}}, y\right):= \mathcal{N}\left(\frac{c_1 \sqrt{\bar{\alpha}_t} {\hat{x}_0^{\prime}}+c_2 x_t}{c_1+c_2}, \frac{c_1c_2}{c_1+c_2} \mathbf{I}\right)
        \label{resample}
    \end{equation}
    where $c_1 = \sigma_t^2$, $c_2 = 1-\bar{\alpha}_t$, and
    \begin{equation}
        \sigma_t^2=\gamma\left(\frac{1-\bar{\alpha}_{t-1}}{\bar{\alpha}_t}\right)\left(1-\frac{\bar{\alpha}_t}{\bar{\alpha}_{t-1}}\right)
        \label{sigma_definition}
    \end{equation}
    following \cite{song2023solving}.
    Larger $\gamma$ indicates a stronger contribution from $\hat{x}_0^{\prime}$.
    The updated state $x_t^{\prime}$ replaces $x_t$, after which the reverse diffusion process resumes.
\end{enumerate}

Data consistency is not applied during early diffusion steps, where samples remain highly noisy and $\hat{x}_0$ may deviate substantially from the underlying solution. 
Instead, enforcement begins from an intermediate time step $t_{\mathrm{start}}$, when diffusion estimates enter the refinement regime and become more sensitive to measurement constraints \cite{yu2023freedom}. 
Furthermore, solving Eq.~\ref{GD} at every time step is computationally redundant. 
Therefore, data consistency is applied periodically (once every 10 steps) after $t_{\mathrm{start}}$, and the corresponding time steps are denoted by $\mathbf{t}_{\mathrm{DC}}$. 
For all time steps, including those not in $\mathbf{t}_{\mathrm{DC}}$, measurement information remains implicitly incorporated through conditioning on $y$ during reverse diffusion.

\section{Experiments}
\label{experiments}
\subsection{Experimental Setup}
\subsubsection{MuPa-ZTE Scanning Settings}
All real-scan MuPa-ZTE datasets were acquired on a GE MR750w scanner (GE HealthCare, Chicago, IL) using a vendor research prototype sequence. 
The acquisition employed 3D center-out radial spokes arranged in phyllotaxis order with an isotropic resolution of $(1.1\,\mathrm{mm})^3$ and a field of view of $(19.2\,\mathrm{cm})^3$. 
The nominal acquisition time was 4 min 33 s.
The pseudo-random phyllotaxis ordering enables retrospective undersampling within the same scan session. 
Accelerated acquisitions were simulated by retaining approximately 25\% of the acquired spokes, corresponding to a prospective scan time of 1 min 9 s.

\subsubsection{Datasets}
For synthetic data, we leveraged BrainWeb \cite{cocosco1997brainweb} digital phantoms to generate anatomy-driven qMRI maps following \cite{sabidussi2021recurrent, wang2024qmri}.
For each brain phantom, five different sets of ground-truth maps that share the same anatomy were generated.
Synthetic weighted images were generated by applying Eq.\ref{Mzspgr}, \ref{Mzi}) together with filp-angle projection to the synthetic qMRI maps followed by k-space encoding and density-compensated reconstruction to reproduce realistic sampling artefacts. 
Accelerated acquisitions were simulated by retaining only the subset of spokes corresponding to the accelerated protocol. 
All encoding and reconstruction operations were implemented using TorchKbNufft\cite{muckley2020torchkbnufft}.
Separate synthetic datasets were created for the nominal and accelerated MuPa-ZTE acquisitions.
For both acquisitions, the training set was created using 95 ground-truth map sets from 19 anatomies, and the test set was created using the remaining 5 sets from a held-out anatomy.

Real scan dataset consists of a NIST/ISMRM phantom, two healthy volunteers, and a patient with brain metastases.
All scans were performed with the nominal MuPa-ZTE.
The accelerated MuPa-ZTE for real data was simulated by retrospectively truncating the acquired k-space rather than a separate prospective scan.
The \textit{in vivo} acquisitions were approved by our institutional review board and informed consent was obtained.
The data was anonymised prior to analysis.

\subsubsection{Model Training and Inference}
The DDPM backbone was implemented as a 3D U-Net with time-step embedding following \cite{diff_unet1, diff_unet2}. 
The network comprised three resolution levels with 128, 256, and 256 channels, respectively. 
During training, 3D patches of size $40^3$ were randomly sampled from the synthetic volumes to form mini-batches (batch size 16). 
To simulate acquisition noise, Gaussian noise was added on the fly to the weighted images of each training sample.
Training was performed for 156k iterations using the Adam optimiser with a learning rate of $1\times10^{-5}$. 
Separate models for nominal and accelerated acquisitions were trained from scratch using identical optimisation settings.

During inference, full 3D volumes were processed using an overlapping patch-based strategy \cite{patch_diff} to accommodate GPU memory constraints (patch size $120^3$, stride 60). 
The data consistency hyperparameters ($t_{start}$ and $\tau$) were determined via grid search on the synthetic test dataset to balance reconstruction accuracy and computational cost, and the selected values were fixed for all subsequent experiments. 
Each gradient-descent optimisation step was terminated after 2000 iterations or earlier once the loss threshold $\tau$ was reached. 
The parameter $\gamma$ in Eq.~\ref{sigma_definition} was fixed to 40.

\subsection{Evaluation Protocol}
For clarity in the experimental comparisons, we denote the dictionary matching method, the purely data-driven diffusion model, and the diffusion model with data consistency as DictMatch, DL-Diffusion, and DL-Diffusion-DC, respectively.
The evaluation protocol was designed to assess quantitative accuracy, robustness under acceleration, and stability across synthetic and real datasets.

Prior to evaluation, the input real scan weighted images for all methods were reconstructed from the same k-space data using density-compensated adjoint non-uniform FFT. 
Because synthetic data were generated from proton-density maps scaled to the range $[0,1]$ (a.u.), whereas real-scan intensities are affected by acquisition-related factors such as coil sensitivity variations, an intensity normalisation step was applied to real data. 
Specifically, a 3D second-order polynomial was fitted to the head region of the PD-weighted image and used as a spatial scaling field for all weighted images, thereby aligning intensity ranges between synthetic and real datasets while compensating for low-frequency intensity inhomogeneity.
For diffusion model-based methods, density compensation proportional to $|k|^2$ was used during weighted-image reconstruction to improve spatial sharpness, with subsequent noise suppression handled by the learned prior. 
For dictionary matching, a gridding-based density compensation with limited compensation factors was adopted to avoid excessive noise amplification, ensuring stable parameter estimation under both nominal and accelerated acquisitions.

The evaluation followed a staged design consisting of four experiments: 1) hyperparameter selection on synthetic data, 2) quantitative validation using a phantom, 3) qualitative and semi-quantitative evaluation on healthy volunteers, and 4) qualitative assessment on patient data.

\subsubsection{Hyperparameter Selection}
Data consistency (DC) hyperparameters were selected using the synthetic test dataset while simultaneously comparing DL-Diffusion-DC with DL-Diffusion. 
We evaluated combinations of $t_{start} \in \{200, 400, 600\}$ (out of a total of $T=1000$ diffusion steps) and $\tau \in \{0.005, 0.003, 0.001, 0.0005\}$ using 30 patches (size $40^3$) sampled from the test set under both nominal and accelerated acquisitions.
Mean absolute error (MAE) with respect to the ground-truth maps and computational time were used as the selection criterion. 
The selected hyperparameters were subsequently fixed and applied to all remaining experiments on real-scan datasets.

\subsubsection{Evaluation on Phantom}
Quantitative evaluation on real data was first performed using the NIST/ISMRM phantom, with reference values specified at 3.0T and 22\textcelsius. 
T1 and T2 mapping accuracy were assessed using the corresponding layers of spheres in the phantom. 
For each method, spheres were segmented from the reconstructed maps, and the mean T1 or T2 values within each sphere were computed and compared with the reference values. 
All phantom experiments were conducted under the nominal MuPa-ZTE acquisition.

\subsubsection{Evaluation on Healthy Volunteers}
For \textit{in vivo} experiments, where ground-truth maps are unavailable, evaluation focused on qualitative assessment and relative quantitative comparison. 
PD, T1, and T2 maps from each method were visually inspected, with particular attention to delineation of fine anatomical structures and the presence of noise or blurring. 
Each method was evaluated under both nominal and accelerated MuPa-ZTE acquisitions. 
For semi-quantitative analysis, DictMatch results from the nominal acquisition were treated as a reference standard for each volunteer. 
Within the brain region, 60 small homogeneous regions of interest (ROIs) of size $5^2$ voxels were placed, 30 in white matter (WM) and 30 in gray matter (GM). 
Mean T1 and T2 values were computed within each ROI for all methods and acquisition settings. 
Absolute differences relative to the reference values were compared, and statistical significance was assessed using a Wilcoxon signed-rank test across ROIs.
Additionally, because diffusion models are stochastic, each diffusion-based method was additionally executed 10 times on one volunteer to assess inference repeatability. 
Voxelwise standard deviations across repeated outputs were computed to produce uncertainty maps for PD, T1, and T2. 
Lower uncertainty, particularly in anatomically relevant regions, indicates improved stability and reproducibility.

\subsubsection{Evaluation on Patient Data}
Finally, the methods were applied to patient data to qualitatively assess robustness and feasibility in the presence of pathology and to examine whether accelerated acquisition remains informative in such scenarios. 
As ground-truth maps were unavailable, evaluation was limited to visual assessment. 
Maps were examined with respect to lesion delineation, boundary clarity, and overall image quality under both nominal and accelerated acquisition settings. 
No diagnostic conclusions were drawn and this experiment serves solely as a qualitative illustration of performance in a clinical-like setting.

\section{Results}
\subsection{Result of Synthetic Data Experiments}
\begin{figure}
    \centering
    \includegraphics[width=\linewidth]{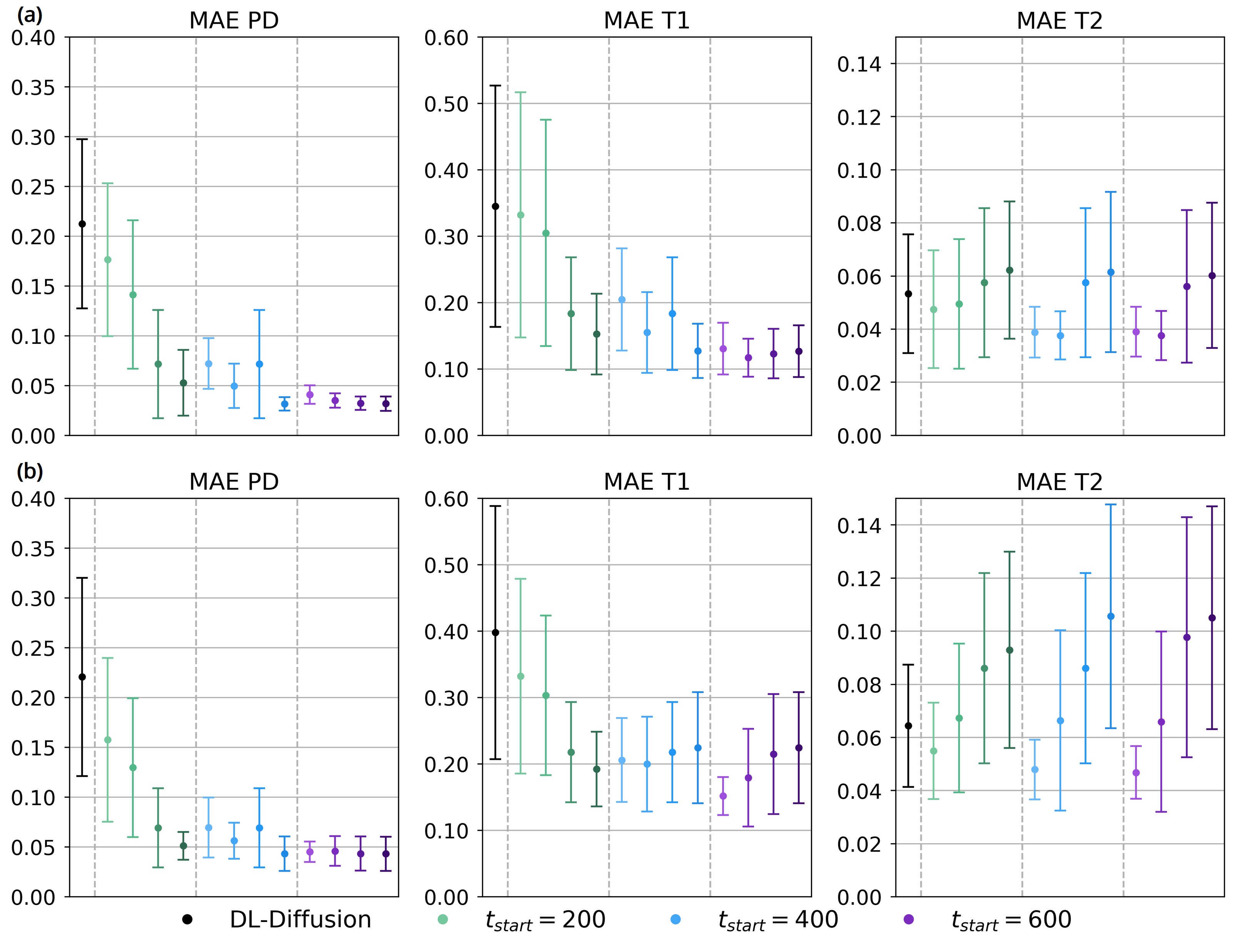}
    \caption{Hyperparameter selection for data consistency (DC) on synthetic datasets. Mean absolute error (MAE) distributions computed over 30 test patches are shown for PD, T1, and T2 mapping under (a) nominal and (b) accelerated MuPa-ZTE acquisitions. Black markers denote the purely data-driven diffusion model (DL-Diffusion), while coloured markers represent DL-Diffusion-DC with different hyperparameter configurations. Colours indicate the starting diffusion step for data consistency ($t_{start}=200, 400, 600$), and colour intensity encodes the optimisation threshold $\tau$ (lighter to darker corresponds to decreasing $\tau\in \{0.005, 0.003, 0.001, 0.0005\}$). Error bars represent mean $\pm$ standard deviation across patches.}  
    \label{fig3}
\end{figure}

Fig.~\ref{fig3} summarises the comparison between DL-Diffusion and DL-Diffusion-DC across different DC hyperparameter configurations on synthetic datasets.
For nominal MuPa-ZTE (Fig.~\ref{fig3}(a)), incorporating data consistency led to consistent improvements in PD and T1 mapping accuracy compared with DL-Diffusion.
In general, increasing $t_{start}$ or decreasing $\tau$ led to lower MAE, although performance gains gradually saturated for very small $\tau$ values.
For T2 mapping, aggressive data consistency (particularly very low $\tau$) occasionally degraded performance, suggesting that excessive optimisation may over-constrain parameters that are less strongly determined by the forward model.
For accelerated MuPa-ZTE (Fig.~\ref{fig3}(b)), PD mapping benefited more noticeably from stronger data consistency.
For T1 mapping, reducing $\tau$ improved performance when $t_{start}=200$, while larger $t_{start}$ values showed less consistent behaviour.
In contrast, T2 estimation was generally more sensitive to DC strength, and several configurations produced slightly higher MAE than DL-Diffusion.
Nevertheless, all DC configurations improved PD and T1 accuracy relative to DL-Diffusion under both acquisition settings.

Although larger $t_{start}$ values and smaller $\tau$ generally yielded lower MAE, these settings substantially increased computational cost because additional optimisation steps must be performed during inference for each patch.
In practice, configurations with $t_{start}=400$ or $600$ increased the total inference time to approximately 10–20 hours per volume, whereas $t_{start}=200$ allowed inference to be completed within approximately 6 hours across all tested $\tau$ values.
Considering the diminishing accuracy gains relative to the increased computational burden, we selected $t_{start}=200$ and $\tau=0.001$ as the operating configuration for DL-Diffusion-DC in all subsequent experiments.
This configuration provided consistent improvements in PD and T1 mapping over DL-Diffusion while preserving T2 performance, achieving a favourable balance between quantitative accuracy and computational efficiency.

\subsection{Results of Evaluation on Phantom}
\begin{figure}
    \centering
    \includegraphics[width=\linewidth]{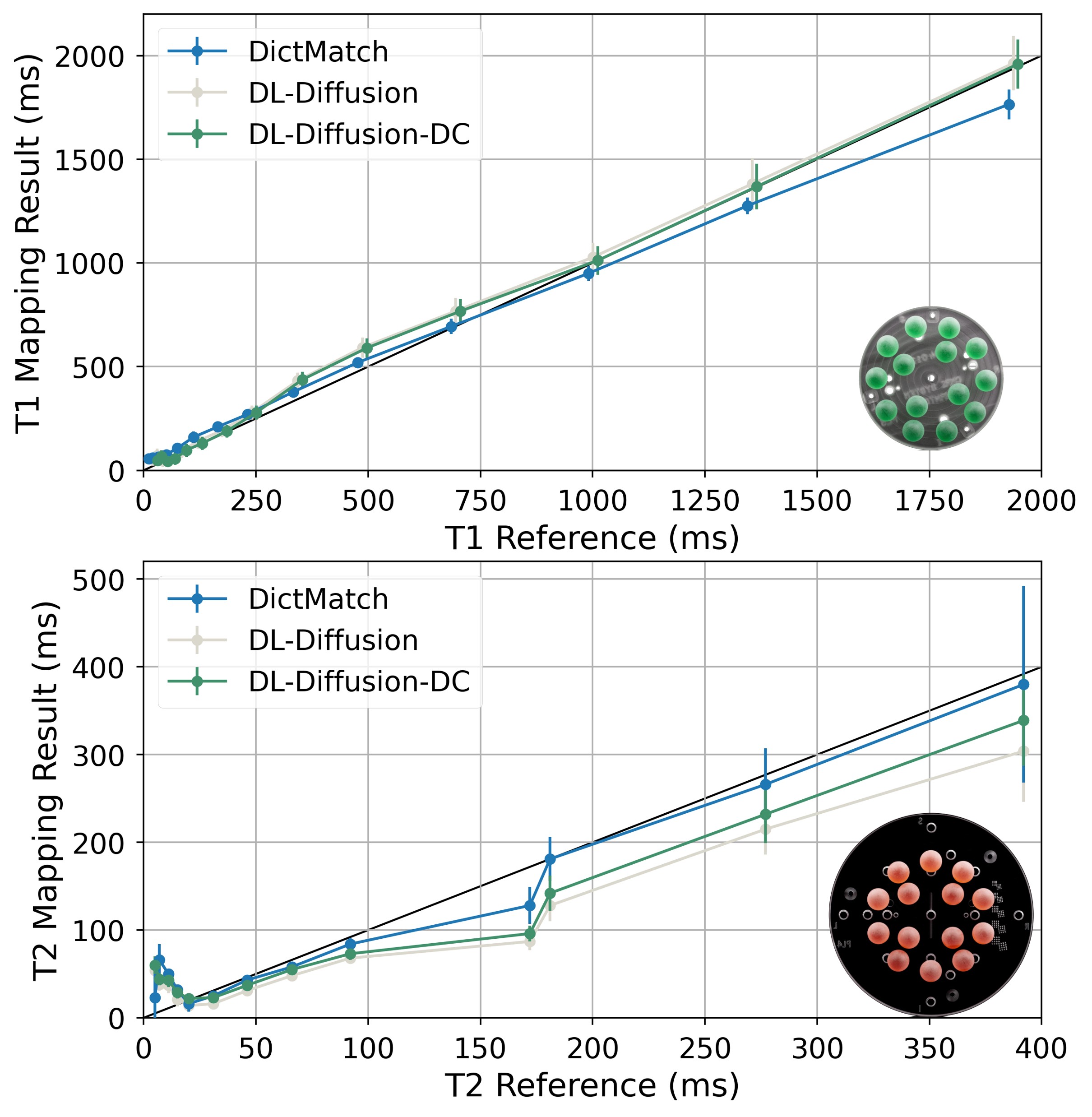}
    \caption{Quantitative evaluation of T1 and T2 mapping using the NIST/ISMRM phantom. All results were produced based on nominal MuPa-ZTE acquisition. Each plotted point corresponds to the mean value within one phantom sphere. Error bars represent $\pm$1 standard deviation. Data points of T1 mapping results are horizontally offset for clarity.}
    \label{fig4}
\end{figure}

The T1 and T2 mapping results obtained from the NIST/ISMRM phantom are shown in Fig.~\ref{fig4}.
For T1 mapping, all three methods achieved high accuracy across most spheres, with estimates closely following the reference values.
Across mid-to-long T1 ranges, DL-Diffusion-DC generally produced estimates closer to the reference line than both DictMatch and DL-Diffusion, although its performance was comparable to DL-Diffusion and slightly inferior to DictMatch for intermediate T1 values (approximately 241–695 ms).
All methods exhibited noticeable bias for the two shortest T1 spheres (21 and 30 ms).
For T2 mapping, DictMatch achieved the highest overall accuracy across most T2 values.
All methods showed bias for very short T2 values ($<$20 ms).
DL-Diffusion-DC improved upon DL-Diffusion but remained less accurate than DictMatch overall.

\subsection{Result of Evaluations on Healthy Volunteer}
\begin{figure}
    \centering
    \includegraphics[width=\linewidth]{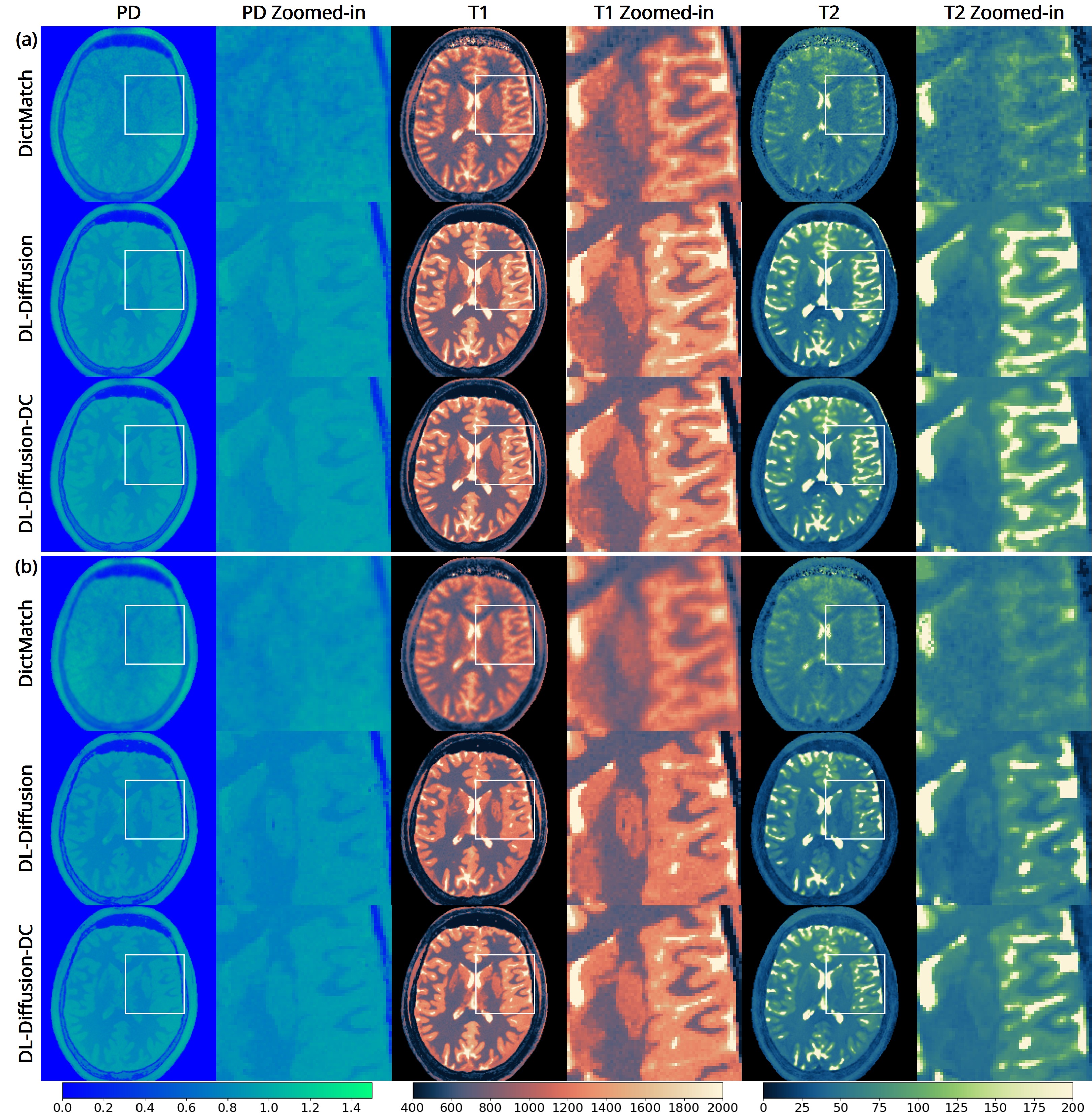}
    \caption{Representative quantitative mapping results (PD in a.u.; T1 and T2 in ms) from a healthy volunteer, comparing DictMatch, DL-Diffusion, and DL-Diffusion-DC under (a) nominal MuPa-ZTE and (b) accelerated MuPa-ZTE acquisitions.}
    \label{fig5}
\end{figure}

Fig.~\ref{fig5} presents representative mapping results from one healthy volunteer.

For nominal MuPa-ZTE (Fig.~\ref{fig5}(a)), diffusion model-based methods exhibit reduced noise appearance and clearer tissue interfaces compared with DictMatch.
In the T1 maps, tissue boundaries appear more distinct in both DL-Diffusion and DL-Diffusion-DC, whereas DictMatch shows increased noise and less clearly defined edges.
In the T2 maps, boundaries between gray matter (GM) and white matter (WM) are less distinguishable in DictMatch but are visually discernible in both diffusion-based methods.
Similarly, PD maps produced by diffusion-based methods show enhanced structural contrast compared with DictMatch.

For accelerated MuPa-ZTE (Fig.~\ref{fig5}(b)), DictMatch exhibits substantial degradation, with blurred structures and reduced tissue contrast.
Both DL-Diffusion and DL-Diffusion-DC recover anatomically coherent structures despite using approximately 25\% of the acquired data.
DL-Diffusion shows noticeable deviations relative to its nominal-acquisition counterpart, including a global shift in GM T1 values.
In comparison, DL-Diffusion-DC produces maps that more closely resemble those obtained under nominal acquisition.
Fine anatomical features, such as the thin cerebrospinal fluid (CSF) layer surrounding the cortex, remain visible in DL-Diffusion-DC but appear less distinct in DL-Diffusion.

For quantitative comparison, Table~\ref{table1} reports the average T1 and T2 values measured in white matter and gray matter ROIs across both volunteers (WM-1, WM-2, GM-1, GM-2).
DictMatch results from nominal MuPa-ZTE were treated as reference values.
A Wilcoxon signed-rank test was used to assess differences relative to the reference.
For nominal MuPa-ZTE, DL-Diffusion-DC showed small deviations that were not statistically significant ($p>0.05$ for all ROIs in both T1 and T2), whereas DL-Diffusion exhibited larger deviations that reached statistical significance across all ROI groups.
Under accelerated MuPa-ZTE, both diffusion-based methods showed increased deviations from the reference.
Nevertheless, DL-Diffusion-DC demonstrated smaller absolute differences than DL-Diffusion for T1 across all ROI groups and for T2 in three of four ROI groups.

\begin{table}
\resizebox{\textwidth}{!}{
\begin{tabular}{clcccccccccc}
\hline
\multicolumn{1}{l}{}                                                       & \multicolumn{1}{c}{} & \multicolumn{3}{c}{\begin{tabular}[c]{@{}c@{}}Mean T1\\ (Difference w.r.t. DictMatch)\end{tabular}}                                         & \multicolumn{2}{c}{\begin{tabular}[c]{@{}c@{}}T1 mapping\\ P-values\end{tabular}} & \multicolumn{3}{c}{\begin{tabular}[c]{@{}c@{}}Mean T2\\ (Difference w.r.t. DictMatch)\end{tabular}}                                          & \multicolumn{2}{c}{\begin{tabular}[c]{@{}c@{}}T2 mapping\\ P-values\end{tabular}} \\ \hline
\multicolumn{1}{l}{}                                                       &                      & DictMatch            & DL-Diff                                                 & DL-Diff-DC                                                          & $P_{DL}$                                   & $P_{DL-DC}$                                   & DictMatch          & DL-Diff                                                       & DL-Diff-DC                                                       & $P_{DL}$                                   & $P_{DL-DC}$                                   \\
\multirow{4}{*}{\begin{tabular}[c]{@{}c@{}}nominal \\ MuPa-ZTE\end{tabular}}  & WM-1                 & \textit{735}  & \begin{tabular}[c]{@{}c@{}}758\\ (+23)\end{tabular}  & \textbf{\begin{tabular}[c]{@{}c@{}}729\\ (-6)\end{tabular}}   & \textless{}0.01                         & 0.40                                    & \textit{53} & \begin{tabular}[c]{@{}c@{}}48\\ (-5)\end{tabular}          & \textbf{\begin{tabular}[c]{@{}c@{}}50\\ (-3)\end{tabular}} & \textless{}0.01                         & \textless{}0.01                         \\
                                                                           & WM-2                 & \textit{695}  & \begin{tabular}[c]{@{}c@{}}720\\ (+25)\end{tabular}  & \textbf{\begin{tabular}[c]{@{}c@{}}690\\ (-5)\end{tabular}}   & \textless{}0.01                         & 0.39                                    & \textit{51} & \begin{tabular}[c]{@{}c@{}}46\\ (-5)\end{tabular}          & \textbf{\begin{tabular}[c]{@{}c@{}}50\\ (-1)\end{tabular}} & \textless{}0.01                         & 0.08                                    \\
                                                                           & GM-1                 & \textit{1249} & \begin{tabular}[c]{@{}c@{}}1303\\ (+54)\end{tabular} & \textbf{\begin{tabular}[c]{@{}c@{}}1269\\ (+20)\end{tabular}} & \textless{}0.01                         & 0.16                                    & \textit{72} & \begin{tabular}[c]{@{}c@{}}83\\ (+11)\end{tabular}         & \begin{tabular}[c]{@{}c@{}}83\\ (+11)\end{tabular}         & \textless{}0.01                         & \textless{}0.01                         \\
                                                                           & GM-2                 & \textit{1058} & \begin{tabular}[c]{@{}c@{}}1109\\ (+51)\end{tabular} & \textbf{\begin{tabular}[c]{@{}c@{}}1071\\ (+13)\end{tabular}} & 0.01                                    & 0.65                                    & \textit{60} & \begin{tabular}[c]{@{}c@{}}73\\ (+13)\end{tabular}         & \begin{tabular}[c]{@{}c@{}}73\\ (+13)\end{tabular}         & \textless{}0.01                         & \textless{}0.01                         \\ \hline
\multirow{4}{*}{\begin{tabular}[c]{@{}c@{}}accelerated \\ MuPa-ZTE\end{tabular}} & WM-1                 & -             & \begin{tabular}[c]{@{}c@{}}702\\ (-33)\end{tabular}  & \textbf{\begin{tabular}[c]{@{}c@{}}720\\ (-15)\end{tabular}}  & \textless{}0.01                         & 0.04                                    & -           & \begin{tabular}[c]{@{}c@{}}42\\ (-11)\end{tabular}         & \textbf{\begin{tabular}[c]{@{}c@{}}49\\ (-4)\end{tabular}} & \textless{}0.01                         & \textless{}0.01                         \\
                                                                           & WM-2                 & -             & \begin{tabular}[c]{@{}c@{}}757\\ (+62)\end{tabular}  & \textbf{\begin{tabular}[c]{@{}c@{}}747\\ (+52)\end{tabular}}  & \textless{}0.01                         & \textless{}0.01                         & -           & \begin{tabular}[c]{@{}c@{}}43\\ (-8)\end{tabular}          & \textbf{\begin{tabular}[c]{@{}c@{}}49\\ (-2)\end{tabular}} & \textless{}0.01                         & 0.09                                    \\
                                                                           & GM-1                 & -             & \begin{tabular}[c]{@{}c@{}}1186\\ (-63)\end{tabular} & \textbf{\begin{tabular}[c]{@{}c@{}}1281\\ (+32)\end{tabular}} & \textless{}0.01                         & 0.03                                    & -           & \begin{tabular}[c]{@{}c@{}}64\\ (-8)\end{tabular}          & \textbf{\begin{tabular}[c]{@{}c@{}}77\\ (+5)\end{tabular}} & \textless{}0.01                         & 0.04                                    \\
                                                                           & GM-2                 & -             & \begin{tabular}[c]{@{}c@{}}1100\\ (+42)\end{tabular} & \textbf{\begin{tabular}[c]{@{}c@{}}1097\\ (+39)\end{tabular}} & 0.09                                    & 0.05                                    & -           & \textbf{\begin{tabular}[c]{@{}c@{}}64\\ (+4)\end{tabular}} & \begin{tabular}[c]{@{}c@{}}69\\ (+9)\end{tabular}          & 0.28                                    & 0.04                                    \\ \hline
\end{tabular}
}
\caption{Semi-quantitative evaluation of T1 and T2 mapping in white matter (WM) and gray matter (GM) ROIs for two healthy volunteers. Values are reported relative to nominal MuPa-ZTE DictMatch results used as reference. Statistical significance was assessed using the Wilcoxon signed-rank test.}
\label{table1}
\end{table}

\begin{figure}
    \centering
    \includegraphics[width=0.75\linewidth]{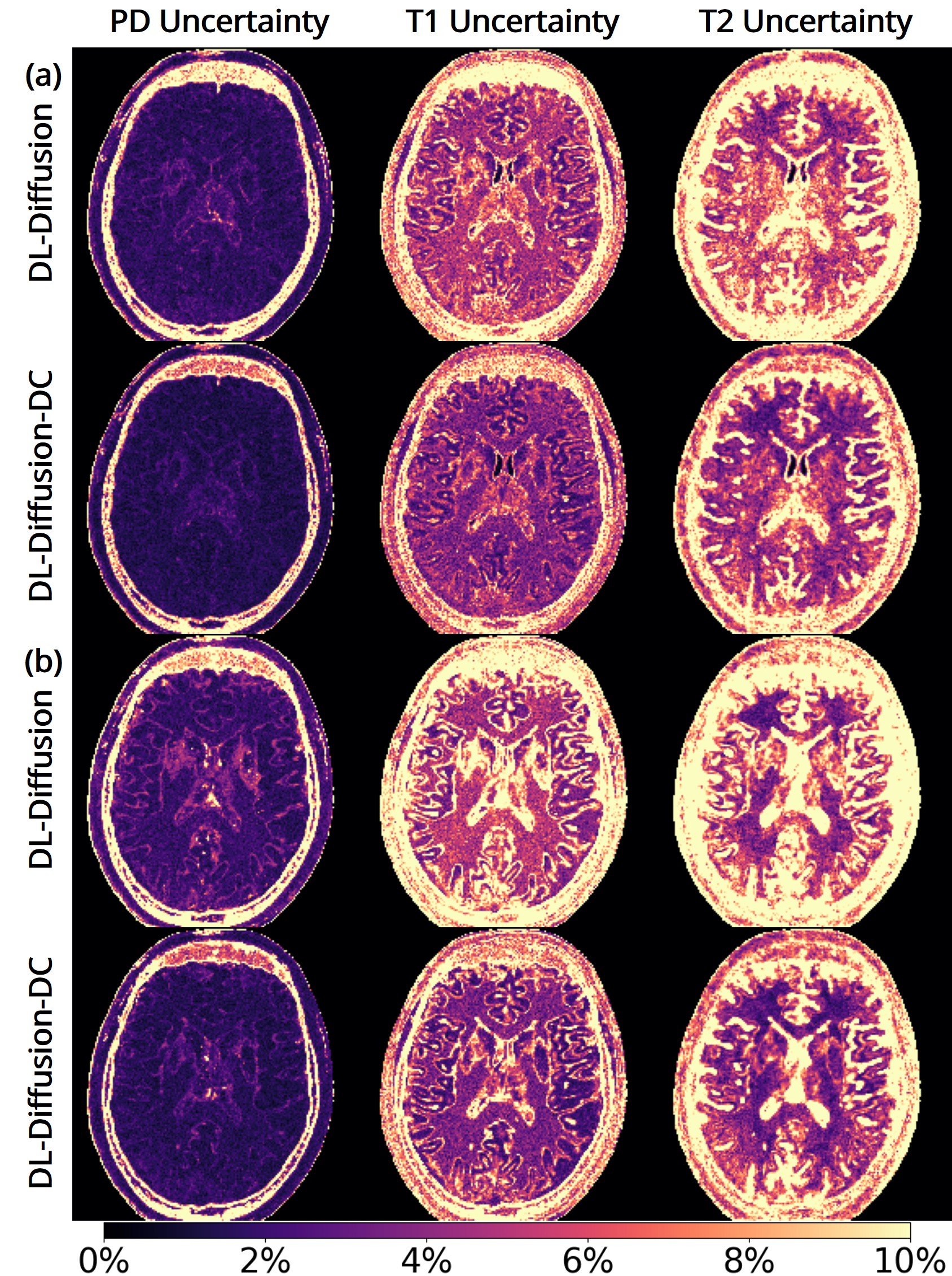}
    \caption{Representative uncertainty maps comparing DL-Diffusion and DL-Diffusion-DC under (a) nominal MuPa-ZTE and (b) accelerated MuPa-ZTE. Values represent the voxel-wise standard deviation across 10 repeated inferences normalised by their mean.}
    \label{fig6}
\end{figure}

Fig.~\ref{fig6} presents representative uncertainty maps, where lower values indicate higher reproducibility across repeated inferences.
For both nominal (Fig.~\ref{fig6}(a)) and accelerated (Fig.~\ref{fig6}(b)) acquisitions, elevated uncertainty is primarily observed in CSF regions and at tissue interfaces.
Accelerated acquisitions exhibit overall higher uncertainty levels.
Compared with DL-Diffusion, DL-Diffusion-DC exhibits lower uncertainty values, particularly in ventricular regions of PD maps and within WM regions in both T1 and T2 maps.
In addition, reduced uncertainty is observed along GM–WM boundaries in T1 maps when data consistency is incorporated.

\subsection{Results of Evaluation on Patient}
\begin{figure}
    \centering
    \includegraphics[width=\linewidth]{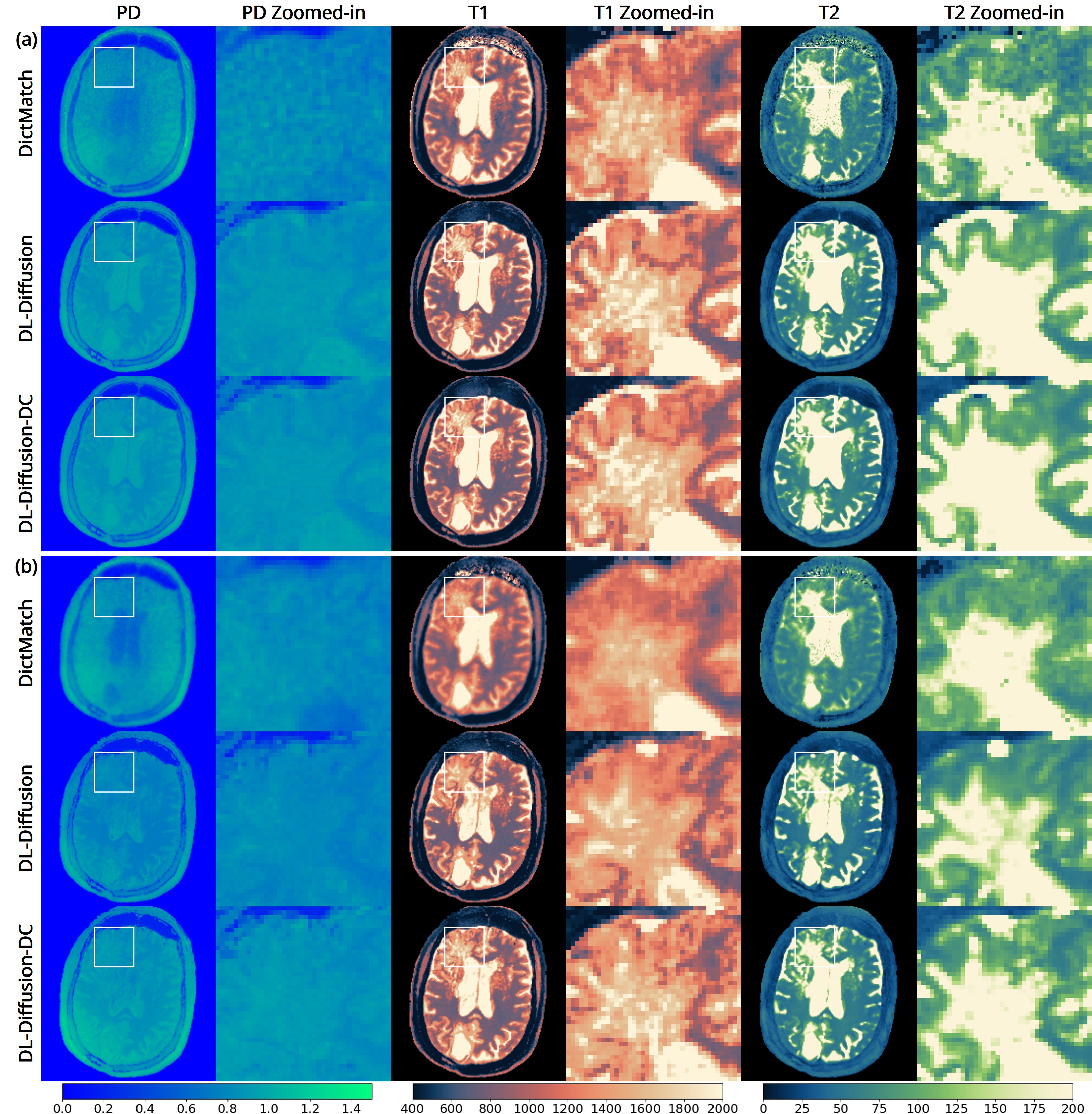}
    \caption{Representative quantitative mapping results (PD in a.u.; T1 and T2 in ms) from the patient dataset, illustrating method performance in the presence of pathology. Results from DictMatch, DL-Diffusion, and DL-Diffusion-DC are shown for (a) nominal MuPa-ZTE and (b) accelerated MuPa-ZTE acquisitions.}
    \label{fig7}
\end{figure}

Fig.~\ref{fig7} presents axial slices of the mapping results for the patient, together with a zoomed-in view of a lesion region.
For nominal MuPa-ZTE (Fig.~\ref{fig7}(a)), both DL-Diffusion and DL-Diffusion-DC show clearer delineation of anatomical structures compared with DictMatch, where image appearance is more affected by noise.
Under accelerated MuPa-ZTE (Fig.~\ref{fig7}(b)), DictMatch results become markedly blurred, while diffusion model–based methods preserve major structural features, although with reduced sharpness compared with the nominal acquisition.
In the zoomed-in lesion region, the T1 map produced by DL-Diffusion exhibits fewer voxels with high T1 values under accelerated acquisition relative to the nominal case, whereas this discrepancy is less pronounced for DL-Diffusion-DC.
Outside the zoomed region, consistent with observations in healthy volunteers, diffusion model–based methods provide improved visual depiction of anatomical structures compared with DictMatch.

\section{Discussion}
This work investigates qMRI mapping for the MuPa-ZTE acquisition and shows that, as acquisition time is reduced to enable clinically practical scan durations (i.e., $\sim$4min or even $\sim$1min), the robustness of the reconstruction stage becomes an increasingly important factor in determining overall qMRI quality.
Our results indicate that conventional model-based mapping approaches become sensitive to noise and undersampling and such limitations are amplified as scanning time is reduced, shifting the dominant practical limitation from acquisition to reconstruction. 
The proposed physics-informed diffusion framework addresses this by combining a learned generative prior with explicit enforcement of acquisition-consistent physics information during inference.

The experiments demonstrate complementary roles of generative modelling and data consistency. 
Purely data-driven diffusion models provide strong structural priors that suppress noise and recover anatomical details, particularly under accelerated acquisition. 
However, without explicit physical constraints, the inferred parameter maps may exhibit bias or reduced fidelity to the measurements. 
Introducing data consistency constrains the reverse diffusion process toward solutions compatible with the MuPa-ZTE forward model, improving quantitative agreement and reducing variability across repeated inferences. 
Across both nominal and accelerated acquisitions, data consistency improved quantitative agreement, with the relative benefit becoming more pronounced under accelerated sampling.
This behaviour suggests that diffusion models effectively explore plausible solution spaces, while physics-based constraints stabilise the inference by restricting it to physically consistent regions.

An important aspect of the proposed approach is its acquisition-aware design. 
Unlike methods that enforce data consistency directly in k-space, we incorporate physics guidance in the weighted-image domain. 
This design choice reflects the complexity of the MuPa-ZTE acquisition, which involves segmented non-Cartesian sampling and multiple magnetisation states, making explicit k-space modelling computationally demanding for 3D diffusion-based inference. 
By applying gradient-based refinement only at selected diffusion steps, the method achieves a practical balance between physical interpretability and computational feasibility while remaining tailored to the acquisition characteristics of MuPa-ZTE.

Another notable observation is that models trained entirely on synthetic data generalised well to phantom, healthy volunteer, and patient datasets. 
This suggests that when synthetic qMRI maps sufficiently resemble realistic tissue characteristics, and when the forward model and acquisition process are adequately represented, synthetic data can provide effective supervision for learning robust priors for qMRI mapping.
In this context, the diffusion model primarily learns the distribution of quantitative parameters together with anatomical and statistical regularities, while the data consistency step anchors the solution to the measured data, thereby mitigating domain discrepancies between synthetic and real acquisitions.

Several limitations should be acknowledged. 
First, T2 estimation remained more challenging than T1 mapping, which is largely attributable to the information encoding characteristics of the MuPa-ZTE sequence, where only limited contrast information encodes T2 relaxation. 
Second, slight expansion of CSF regions was occasionally observed, likely reflecting biases inherited from the synthetic training data. 
Third, validation was performed on a limited number of subjects, and broader evaluation across diverse pathologies would be required to fully assess clinical robustness. 
Finally, the selection of data consistency hyperparameters relied on a restricted search space and empirical criteria. 
Future work may investigate adaptive or theoretically guided strategies for parameter selection.

Overall, this work demonstrates that robust quantitative mapping under fast multi-parametric acquisition benefits from jointly considering acquisition characteristics and reconstruction methodology. 
By integrating MuPa-ZTE with an acquisition-consistent diffusion-based reconstruction strategy, the proposed q3-MuPa framework enables stable 3D qMRI mapping across both nominal and substantially accelerated scans in our experiments.
More broadly, the results suggest that as qMRI acquisitions become increasingly efficient, reconstruction stability increasingly determines the achievable quality of qMRI mapping. 
In this context, incorporating known forward models into generative diffusion processes provides a practical pathway toward reliable and computationally feasible quantitative imaging, enabling fast, high-resolution multi-parametric acquisition that can serve as a quantitative reference within clinical protocols.

\section*{Acknowledgements}
This work was performed within the “Trustworthy AI for MRI” ICAI Lab and was supported by the Dutch Research Council (NWO) and GE HealthCare under project number KICH3.LTP.20.006.




\printcredits

\bibliographystyle{cas-model2-names}

\bibliography{refs}



\end{document}